\documentclass{ajour}

\usepackage{amsmath, epsfig}

\begin{document}

\title{General Relativistic Stars: Linear Equations of State}
\author{Ulf S. Nilsson}
\affil{Department of Applied Mathematics\\University of
  Waterloo\\Waterloo, Ontario\\Canada, N2L 3G1\\and\\
Department of Physics\\
Stockholm University\\
Box 6730\\
S-113 85 Stockholm\\
Sweden}
\email{unilsson@math.uwaterloo.ca}
\and
\author{Claes Uggla}
\affil{Department of Physics
  \\University of Karlstad\\S-651 88 Karlstad\\Sweden} 
\email{uggla@physto.se}

\authorrunninghead{U. S. Nilsson and C. Uggla}
\titlerunninghead{General Relativistic Stars}

\abstract{In this paper Einstein's field equations, for static
  spherically symmetric perfect fluid models with a linear barotropic
  equation of state,  
  are recast into a 3-dimensional {\em regular} system of ordinary 
  differential equations on a compact state space. The system is 
  analyzed qualitatively, using the theory of dynamical systems, and
  numerically. It is shown that certain special solutions play
  important roles as building blocks for the solution
  structure in general. In particular, these special solutions 
  determine many of the features exhibited by solutions with a regular
  center and large central pressure. It is also shown that the present
  approach can be applied to more general classes of barotropic
  equations of state.} 






 
\begin{article}

  
\section{Introduction}
In both Newtonian gravity and general relativity, the simplest
models of isolated stars are given by static spherically symmetric
configurations. Despite their simplicity, they are believed to yield many
insights about much wider classes of stellar models 
(see, for example, the discussion by Hartle
\cite{art:Hartle1978}). 

The line element for a static spherically symmetric model can be
written as 
\begin{equation}
\label{eq:ds2}
 ds^2 = -{\rm e}^{2\phi(\lambda)}dt^2 +
 r(\lambda)^2\left[\tilde{N}(\lambda)^2d\lambda^2 + d\Omega^2\right]\ ,
\end{equation}
with
\begin{equation}
  d\Omega^2 = d\theta^2 + \sin^2\theta d\varphi^2\ ,
\end{equation}
where $\phi(\lambda)$ is the gravitational potential, $\tilde{N}(\lambda)$
a dimensionless (under scale-transformations)
freely specifiable function, and $r(\lambda)$ the usual
Schwarzschild radial parameter, associated with the area of the
spherical symmetry surfaces. The coordinate $\lambda$ is a spatial
radial variable, whose interpretation depends on the
choice of $\tilde{N}$. Since $\tilde{N}=1$
corresponds to isotropic coordinates, the function $\tilde{N}$ can
be viewed as a relative gauge function with respect to the isotropic
gauge. 

This paper is the first in a series devoted to the study of general 
relativistic star models with perfect fluid sources. Thus the
energy-momentum tensor is assumed to be of the form
\begin{equation}
 \label{eq:emom}
 T_{ab} = \rho u_a u_b + p\left(g_{ab} + u_a u_b \right)\ ,
\end{equation}
where $\rho$ is the energy density, $p$ the pressure, and
$u^a$ the 4-velocity of the fluid. A relation between the
gravitational potential $\phi$ in (\ref{eq:ds2}) and the matter 
content can be found from 
the equations of motions for the fluid, $\nabla_a T^{ab}=0$, namely
\begin{equation}
 \label{eq:dphidp} 
 \frac{d\phi}{dp} = -\frac{1}{\rho+p}\ .
\end{equation}
This equation is the general relativistic generalization of the
corresponding Newtonian equation $d\phi/dp = -1/\rho$. 

For a static spherically symmetric perfect fluid model to be
considered as describing a star, we require that it is 
isolated in the sense that the model has a boundary at a finite radius
where the pressure of the fluid vanishes. At this radius, the 
interior solution is matched with a static exterior vacuum spacetime, 
described by the Schwarzschild solution
\begin{equation}\label{eq:schwa}
  ds^2 = -\left(1-\tfrac{2M}{r}\right)dt^2 +
  \frac{dr^2}{\left(1-\tfrac{2M}{r}\right)} + r^2d\Omega^2\ ,
\end{equation}
(see Schwarzschild \cite{art:Schwarzschild1916}). 
A vanishing pressure is, in fact, a
necessary and sufficient matching condition (see, for example,
Stephani \cite{book:Stephani1982}, p 161). 

In order to analyze the
gravitational field equations for (\ref{eq:ds2}), an equation of state
must be specified. Over the years, a number of exact
static spherically symmetric perfect fluid solutions have been found,
see, for example, Delgaty \& Lake \cite{art:DelgatyLake1998} for a
comprehensive review. Unfortunately, most equations of
state for known exact solutions have no physical motivation. They 
are instead chosen solely with the purpose of simplifying the
differential equations, and thereby allowing exact solutions to be 
found. Our aim is not to find new exact solutions, but to gain an
understanding of the solution space, and its implications, for given
equations of state which can be physically motivated. 

We will consider barotropic equations of state
\begin{equation}
  \label{eq:barotate}
  \rho = \rho(p)\ ,
\end{equation}
which dominate in the literature
(see Stergioulas \cite{art:stergioulas} and references therein). 
Such equations of state are relevant
for describing neutron stars and white dwarfs, see, 
for example, Misner {\em et al} \cite{book:MTW1973}, p 624. As a first
step we will consider complete analytic equations of state covering
the pressure range of the entire star models. Once the solution space
for such equations of states are understood, models with composite
equations of state ({\em i.e.} models where the equations of state are
obtained by matching different equations of state in different
pressure regions) can be obtained by matching different matter solutions
(or in the terminology introduced below, by matching orbits in different state
spaces).  

One might think that static spherically symmetric perfect fluid models
lead to simple problems and that there is not much to be discovered in this 
area of research. In this series of papers we will show that this is not 
the case. Some problems turn out to be quite complicated and one can obtain
new insights.

The most favored approach for studying static stars,
the so-called Tolman--Oppenheimer--Volkoff approach,  
uses the pressure and mass as dependent variables and the Schwarzschild
radial coordinate as the independent variable
(see, for example, section 6.2 in Wald \cite{book:Wald1984}). The
resulting equations, however, are not regular at the center of the
star, and the problem of proving 
existence and uniqueness of solutions requires a quite technical
analysis of singular differential equations, see Rendall \& Schmidt
\cite{art:RendallSchmidt1991}. 

Our method is to recast the field equations
for a given equation of state as a regular 3-dimensional system of ordinary 
differential equations on a compact state space. 
This {\em regularization} means that the problems associated with
the singular nature of the Tolman--Oppenheimer--Volkoff equations are
circumvented. In addition, recasting the field equations into
regularized form on a compact state space allows us to conveniently 
investigate the solution space using powerful methods from 
the theory of dynamical systems. Such methods have been used in
spatially  
homogeneous cosmology with great success, see Wainwright \&
Ellis \cite{book:WainwrightEllis1997} and references therein. The
regularized form is also suitable for numerical calculations. 
The fact that the dynamical system is compact and 3-dimensional is of
great advantage, since this makes it possible to {\em visualize} the
state space and thereby obtain a clear picture of the structure of
the {\em entire} solution space. 

We will show that there exist special solutions that play important 
roles as building blocks for the remaining
solution structure. Some of these solutions 
turn out to determine many of the features exhibited by solutions with a 
regular center and a large central pressure. 
We will also show that the behavior of the solutions at small 
(and to some extent also large) radii is intimately connected with
asymptotic self-similarity, even for non-scale-invariant equations of state.
A useful formulation for the gravitational field equations of the
static spherically symmetric  
perfect fluid models and a good understanding of the corresponding solution
structure may also serve as a starting point for exploring their role in a
broader context, as will be discussed in the concluding remarks.

In this paper we consider the linear equation of state
\begin{equation}
  \label{eq:lineqstate}
  \rho = \rho_0 + (\eta - 1)p\ ,
\end{equation}
where the constants $\rho_0$ and $\eta$ satisfy $\rho_0 \geq 0,\eta
\geq 1$. 
The case $\eta=1$ corresponds to an incompressible fluid with
constant energy density, while the case $\rho_0=0$ describes a
scale-invariant equation of state. The scale-invariant case
has been investigated previously by Collins \cite{art:Collins1985}
using dynamical systems theory. Note that $\eta<2$ corresponds to
non-causal fluids, in which the velocity of sound is greater than that
of light. 

The outline of the paper is as follows: in Section \ref{sec:dynsyst}
we write the gravitational field equations as a 3-dimensional
regular dynamical system on a compact state space
and perform a local analysis of the
resulting equations. We also present a monotone function and describe the 
dynamical behavior on the boundaries. 
In Section \ref{sec:regular} we focus on the regular solutions, and
discuss their behavior. In Section \ref{sec:general} we discuss how
one can adapt the approach of this paper to facilitate studies of more
general barotropic equations of state. We conclude with some remarks in Section
\ref{sec:conclude}.

\vspace{0.7cm}

Throughout the paper, geometric units with $c=G=1$ are used, where $c$
is the speed of light and $G$ the gravitational constant. Roman
indices, $a,b,...=0,1,2,3$ denote spacetime indices.

\section{The dynamical system formulation}
\label{sec:dynsyst}

In order to recast the gravitational field equations for
static spherically symmetric prefect fluid models with a linear
equation of state into a regular dynamical system
on a compact state space, we proceed as follows. 
We first write the line element in the form
\begin{equation}
 ds^2 = -{\rm e}^{2\phi}dt^2 +
 d\ell^2 + e^{2\psi - 2\phi} d\Omega^2\ .
\end{equation}
Then we introduce the variables
\begin{equation}
  \theta = \dot{\psi}\ ,\quad \sigma = \dot{\theta}\ ,\quad
  B = e^{\phi -\psi}\ , 
\end{equation}
where a dot denotes differentiation with respect to $\ell$. The
gravitational field equations, expressed in these variables, are
\begin{eqnarray}
  \dot{\theta} &=& -2\theta^2 + \theta\sigma + B^2 + 16{\pi}p\ ,
  \mathletter{a} \label{eq:theta}\\ 
  \dot{\sigma} &=& -2\theta\sigma + \sigma^2 + 4{\pi}(\rho + 3p)\
  , \mathletter{b}\\ 
  \dot{B} &=& (-\theta + \sigma)B\ , \label{eq:B} \mathletter{c}\\
  8{\pi}p &=& \theta^2 - \sigma^2 - B^2\ . \label{eq:pconstr}\mathletter{d}
\end{eqnarray}
This system is very similar to those obtained in
spatially homogeneous cosmology (see Wainwright \& Ellis
\cite{book:WainwrightEllis1997}), even though the physical
interpretation is quite different. One can hence
import ideas from treatments of spatially homogeneous  
models to the present context. 

To obtain a compact state space we introduce the variables
\begin{equation}
  \label{eq:QSC}
  \left\{Q \ , S\ , C \right\}\ ,
\end{equation}
according to
  \begin{eqnarray}
  Q &=& \frac{\theta}{\sqrt{\theta^2 + 8{\pi}\rho_0/\eta}}\ , \quad
  S = \frac{\sigma}{\sqrt{\theta^2 + 8{\pi}\rho_0/\eta}}\ , \nonumber \\
  C &=& \frac{B^2}{\theta^2 + 8{\pi}\rho_0/\eta}\ .
  \end{eqnarray}
These variables are closely related to those used by Uggla {\em et al} 
\cite{art:UgglaSurmulen1990} in cosmology and by Nilsson {\em et al}
\cite{art:Nilssonetal1998sta} for studying static cylinders with a
linear equation of state. We also introduce
and a new independent variable $\lambda$, defined by
\begin{equation}
 \frac{d\ell}{d\lambda} = \frac{1}{\sqrt{\theta^2 +
 8{\pi}\rho_0/\eta}}\ .
\end{equation}
The above choices correspond to
\begin{equation}
  \label{eq:linchoice}
 \tilde{N}^2 = C\ , \quad {\phi}' = S\ , \quad r^2 =
 \frac{\eta(1-Q^2)}{8\pi\rho_0 C} \ ,
\end{equation}
in (\ref{eq:ds2}), and the prime denotes differentiation with respect
to the independent spatial variable $\lambda$. From
(\ref{eq:linchoice}) it 
follows that $C$ and $1 - Q^2$ are positive.
Integrating (\ref{eq:dphidp}), yields 
\begin{equation}
 e^\phi = \alpha\left( \frac{1 - Q^2}{1 - S^2 - C}
 \right)^{1/\eta}\ ,
\end{equation}
where $\alpha$ is a freely specifiable constant corresponding to the
freedom of scaling the time coordinate $t$ in the line element
(\ref{eq:ds2}). This, in turn, reflects the freedom in specifying the
value of the gravitational potential $\phi$ at some particular value
of $r$. Matching an interior solution with the exterior Schwarzschild
solution, however, fixes this constant. 
For the purpose of interpreting the variables $Q$ and $S$, it 
is worth noting the relation 
\begin{equation}
\frac{d\phi}{d\ln r} = \frac{S}{Q-S}\ .
\end{equation}

In terms of the new variables, the gravitational field equations 
(\ref{eq:theta})-(\ref{eq:B}) takes the form
\begin{eqnarray}
  Q' &=& (1-Q^2)\left(QS - C - 2S^2\right)\ ,
  \mathletter{a}\label{eq:lineqa} \\ 
  S' &=& \tfrac12W\left(2 + \eta - 2QS\right) -
  (1-S^2)\left(1-Q^2+QS\right)\ , \mathletter{b}\label{eq:lineqb} \\
  C' &=& 2\left[ S(1-Q^2) + Q\left(S^2 - W\right)\right]C\
  .\mathletter{c}\label{eq:lineqc}  
\end{eqnarray}
Equation (\ref{eq:pconstr}) leads to
\begin{equation} \label{eq:Wdef}
  W = 1-S^2-C\ ,
\end{equation}
where $W$ is defined by 
\begin{equation}
 W = 8{\pi}\eta^{-1}Cr^2(p + \rho)\ ,
\end{equation}
and satisfies $W\geq0$ if we assume that the weak energy condition
$p + \rho \geq 0$ is satisfied. Equation (\ref{eq:Wdef})
is used to eliminate $W$ in (\ref{eq:lineqa})-(\ref{eq:lineqc}).
It follows from (\ref{eq:Wdef}),
(\ref{eq:lineqb}), and (\ref{eq:lineqc}), that     
\begin{equation}
 W' = \left[ 4QS^2 - 2S Q^2 - \eta S + 2QC\right]W\ .
\end{equation}
Hence, $W=0$ is an invariant subset, as are $Q=\pm1$ and $C=0$, which
is easily seen from (\ref{eq:lineqa}) and
(\ref{eq:lineqc}). These invariant subsets constitute the boundary of the 
physical state space for spherically symmetric models with a 
non-scale-invariant linear equation of state,
{\em i.e.}, they describe the boundary of the part of state space where 
$C>0$, $W>0$ and $\rho_0\neq0$. 
We now include these boundaries and obtain a regular dynamical system
(\ref{eq:lineqa})-(\ref{eq:lineqc}) on a compact state space. 

\begin{figure}[ht]
  \begin{center}
    \epsfig{figure=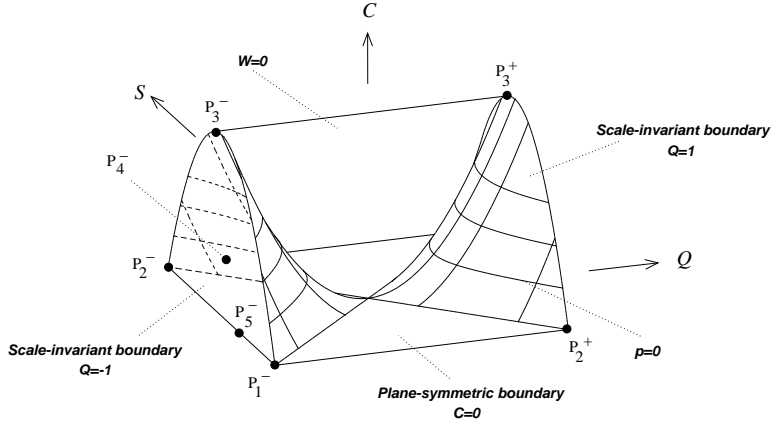, width=0.9\textwidth}
    \caption{The state space for static spherically symmetric models
      with a linear equation of state. The different boundaries are
      identified along with the surface of vanishing pressure, $p=0$.}
    \label{fig:cylstatespace}
  \end{center}
\end{figure}

If the variables (\ref{eq:QSC}) are used for
plane-symmetric models,\footnote{These models have the same line
  element as that in (\ref{eq:ds2}) but $d\Omega^2 = dx^2 + dy^2$.}
(\ref{eq:lineqc}) decouples and the remaining equations are
identical to the $C=0$ subset of (\ref{eq:lineqa})-(\ref{eq:lineqc}). 
It is therefore
natural to refer to this boundary as the plane-symmetric boundary. The
$Q = \pm 1$ subsets correspond to setting $\rho_0 = 0$ 
in (\ref{eq:lineqstate}), and the remaining equations describe models 
with a scale-invariant equation of state $\rho = (\eta
-1)p$.\footnote{To obtain the line element for this case, an additional  
dimensional variable needs to be considered as well.} 
We refer to these boundaries as the scale-invariant boundaries.
The state space, with the different boundaries identified, is
shown in figure \ref{fig:cylstatespace}. 

An important relation is   
\begin{equation}
  \label{eq:pdeflin}
 \frac{\eta p}{\rho_0} = \frac{Q^2 - S^2 - C}{1 - Q^2}\ ,
\end{equation}
since it yields 
\begin{equation}
  \label{eq:psurfacedef}
 Q^2 - S^2 - C \geq 0\ ,
\end{equation}
whenever the pressure is non-negative. 
The expression $Q^2-S^2-C=0$ defines a surface in state space, which we refer
to as the surface of vanishing pressure. This surface is also shown in figure 
\ref{fig:cylstatespace}. It is not an invariant subset of 
(\ref{eq:lineqa})-(\ref{eq:lineqc}).

A monotone function excludes equilibrium points, periodic orbits,
recurrent orbits, and homoclinic orbits in its domain. The function 
\begin{equation}
  Z = \frac{2Q-S}{\sqrt{ (2Q-S)^2 + 3(1-Q^2)}}\ , 
\end{equation}
which satisfies
\begin{equation}
  Z' = -\frac{3(1-Q^2)\left[ 2C + {\eta}W +
  2(2S-Q)^2\right]}{2\left[ (2Q-S)^2 + 3(1-Q^2)\right]^{3/2}}\ , 
\end{equation}
is monotonically decreasing and is defined everywhere in state space
except on the scale-invariant boundaries $Q=\pm1$. Hence, the  
``past'' ($r\rightarrow 0$) and ``future'' ($r\rightarrow+\infty$)
attractors lie on the $Q=\pm1$ boundary subsets. 
On these boundaries, however, there exist other monotone functions
which imply that the only attractors are equilibrium points on the
$Q=\pm1$ boundaries.\footnote{The equations on the 
  boundary subsets $Q=\pm1$ describe 
  models with a scale-invariant equation of state. These equations exhibit
  a monotone function given in Goliath {\em et al} 
  \cite{art:Goliathetal1998b}.} 
The equilibrium points of (\ref{eq:lineqa})-(\ref{eq:lineqc}), 
together with their
corresponding eigenvalues are listed in Table
\ref{tab:cylindrical}. The equilibrium points describe self-similar
exact solutions to the gravitational field equations.
The points  
$P_{1,2}^\pm$ correspond to the plane-symmetric vacuum solutions, first 
found by Kasner \cite{art:Kasner1925}. The points $P_3^\pm$ correspond
to the Minkowski spacetime on explicitly spherically symmetric form. The
points $P_4^\pm$ correspond to a non-regular self-similar perfect fluid
solution due to Tolman \cite{art:Tolman1939}. This solution is, however, also
associated with many other authors, for example, Misner \& Zapolsky
\cite{art:MisnerZapolsky1964}. The points
$P_4^\pm$ change from focuses to nodes at $\eta = \sqrt{64/7} - 2$.  
The equilibrium points $P_5^\pm$
correspond to a self-similar plane-symmetric perfect fluid model. Since
this model only exists for $\eta<2$, the fluid is necessarily non-causal.
There is thus a bifurcation associated with the transition from a 
non-causal to a causal fluid at $\eta=2$. The equilibrium points
$P_5^\pm$ leave the state space through the points $P_1^\pm$
respectively, thereby changing these points from nodes to saddles.

\begin{table}[ht]
  \begin{center}
\begin{tabular}{ccccc}
  \hline
  Eq point & $Q$ & $S$ & $C$ & Eigenvalues\\
  \hline
  $P_1^\pm$ & $\pm 1$ & $\pm1$ & 0 & $\pm(2-\eta)\ , \ \pm2\ , \ \pm2$
   \\ 
  $P_2^\pm$ & $\pm 1$ & $\mp1$ & 0 & $\pm(6+\eta)\ , \ \pm6 \ , \
  \pm2$ \\  
  $P_3^\pm$ & $\pm 1$ & 0 & 1 & $\pm2\ , \ \mp1 \ , \ \pm2$ \\
  $P_4^\pm$ & $\pm 1$ & $\pm\tfrac{2}{2+\eta}$ &
  $\tfrac{a}{(2+\eta)^2}$ &
  $\mp\tfrac{1}{2+\eta}\left[2+\eta 
  +\epsilon\sqrt{ (2+\eta)^2 - 8a } \right] \ , \
  \pm\tfrac{2\eta}{2+\eta}$ \\
  $P_5^\pm$ & $\pm1$ & $\pm\tfrac{1}{4}(2+\eta)$ & 0 &
  $\pm\tfrac{1}{4}(2+\eta)\ , \ \mp\tfrac{1}{8}(2-\eta)(6+\eta) \ , \
  \pm\tfrac{1}{4}a$ \\
  \hline
\end{tabular}
\end{center}
\caption{Equilibrium points and their stability for the linear
  equation of state using the variables $\left\{Q,S,C\right\}$. The
  constant $a$ is given by $a=\eta^2+4\eta-4$. The points $P_5^\pm$
  only exist for $1\leq\eta<2$. The constant $\epsilon$ takes the
  discrete values $\pm1$.}   
    \label{tab:cylindrical}
\end{table}

The line element expressed in the variables (\ref{eq:QSC}) is
invariant under the discrete symmetry
\begin{equation}
  \left(Q,S,\lambda\right) \rightarrow -\left(Q,S,\lambda\right)\ ,
\end{equation}
and since (\ref{eq:lineqa})-(\ref{eq:lineqc}) are also invariant
under this discrete symmetry, a solution is represented
by two orbits in the state space. We can, however, without loss of
generality, focus on orbits entering the state space from the $Q=1$
boundary subset.   

We note that all orbits which correspond to solutions with $\rho_0 \neq 0$,
with spherical as well as plane symmetry, 
start at equilibrium points on the $Q=1$ boundary and end at equilibrium
points on the $Q=-1$ boundary. All solutions are thus asymptotically 
self-similar. However, on their way to $Q=-1$ from $Q=1$ they all
intersect the surface of vanishing pressure, $Q^2-S^2-C=0$, at an interior 
point $(Q_*, S_*,C_*)$ of the state space. 
To obtain physically reasonable spherically symmetric models (models with
non-negative pressure) one matches each interior solution with the exterior 
Schwarzschild vacuum solution at the radius, $R$, where the pressure becomes 
zero. The radius, $R$,
is determined by inserting $(Q_*, S_*, C_*)$ into the expression for
$r$ in (\ref{eq:linchoice}). This leads to
\begin{equation}
  \label{eq:linRexp}
  R = \sqrt{\frac{\eta(1-Q_*^2)}{8\pi\rho_0 (Q_*^2 - S_*^2)}}\ .
\end{equation}
From $Q_*^2-S_*^2-C_*=0$ and $C_*>0$, it follows that $Q_*^2 \neq S_*^2$.
Hence all spherically symmetric solutions are finite.

\subsection{The boundary structure}
\label{sec:boundaries}

\begin{figure}[ht]
  \begin{center}
    \epsfig{figure=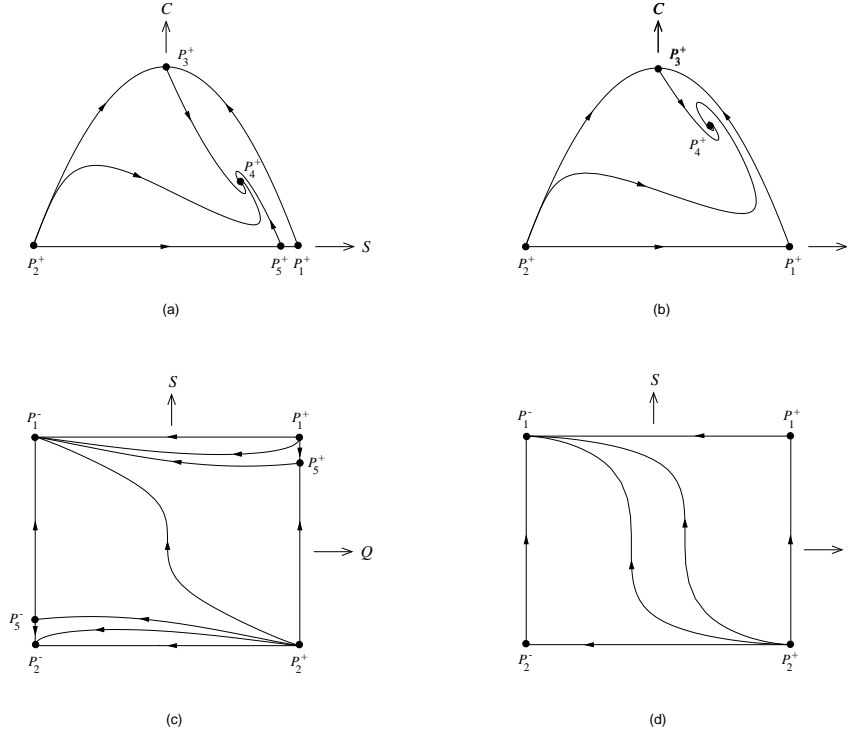, width=\textwidth}
    \caption{Orbits belonging to the boundary subsets for static spherically
      symmetric models with a
      linear equation of state, using the variables $\left\{ Q,S,C
      \right\}$, with (a) the scale-invariant boundary $Q=1$ with
      $1\leq\eta\leq\sqrt{64/7}-2$, (b) the scale-invariant boundary
      $Q=1$ with $\eta>2$, (c) the plane-symmetric boundary $C=0$ with
      $\eta<2$, and (d) the plane-symmetric boundary $C=0$ with
      $\eta\geq2$.}
    \label{fig:etaboundary}
  \end{center}
\end{figure}

To understand the structure of the interior state space, one has
to understand the structure of the boundary subsets.
Orbits belonging to the scale-invariant ($Q=1$) boundary are shown in figure
\ref{fig:etaboundary}a,b while orbits belonging to the plane-symmetric
($C=0$) boundary are shown in figure \ref{fig:etaboundary}c,d. We
refrain 
from explicitly showing orbits for models with $\eta$ in the
interval $\sqrt{64/7} - 2 < \eta < 2$, since these models only require
a slight modification of figure \ref{fig:etaboundary}a.
In addition, the incompressible
fluid case $\eta = 1$ (which belongs to the 
shown $1\leq\eta\leq\sqrt{64/7}-2$ interval), and the
causal fluid cases $\eta \geq 2$, are the most
interesting ones. We require the solutions to have
non-negative pressure and, as seen from the above discussion,
the surface of vanishing pressure cut all solutions before they come
into the neighborhood of the $W=0$ boundary. Thus this boundary plays
a physically less important role than the other boundary subsets and
we therefore refrain from showing orbits belonging to this subset.


\section{Regular solutions}
\label{sec:regular}

Rather than considering all possible orbits, and the associated solutions'
physical features, we focus on orbits corresponding to solutions
with regular centers and positive pressure, the so-called
{\em regular subset} of solutions. 

The spacetime for a regular star is described by the
flat Minkowski geometry at the center.
Hence, all orbits belonging to the regular subset start
from the equilibrium point $P_3^+$, which corresponds to Minkowski
spacetime on spherically symmetric form. In the vicinity of
$P_3^+$, one finds the following approximate expression for the regular
orbits\footnote{These expressions are found from the   
eigenvectors associated with the two positive eigenvalues (both equal to 2)
of the equilibrium point $P_3^+$.}
\begin{eqnarray}
  \label{eq:regsollin}
  Q &=& 1-\epsilon_1{\rm e}^{2\lambda}\ , \mathletter{a}\\
  S &=& \tfrac{2}{3}\left(\epsilon_2-\epsilon_1\right){\rm
  e}^{2\lambda}\ , \mathletter{b}\\
  C &=& 1-\tfrac{4}{2+\eta}\epsilon_2{\rm e}^{2\lambda}\ , \mathletter{c}
\end{eqnarray}
where $\epsilon_1$ and $\epsilon_2$ are small positive constants. It
is, however, only the quotient $0\leq\epsilon_1/\epsilon_2\leq 
+\infty$ that parameterizes the 1-parameter subset of regular
solutions, which can be seen from
\begin{equation}
  \label{eq:epsilonquot}
  \lim_{\lambda\rightarrow -\infty} \frac{p}{\rho} =
  \frac{p_c}{\rho_c} = \frac{\rho_c - \rho_0}{(\eta-1)\rho_c} = \frac{2 -
  (2+\eta)\left(\epsilon_1/\epsilon_2\right)}{2(\eta-1) +
  (2+\eta)\left(\epsilon_1/\epsilon_2\right)}\ ,
\end{equation}
where $p_c$ and $\rho_c$ denote the values of the pressure and energy
density at the center of the star. From (\ref{eq:epsilonquot}) we see
that there exists solutions with a regular center but negative
pressure. For solutions with non-negative pressure at the center, the
quotient 
$\epsilon_1/\epsilon_2$ is subject to the constraint 
\begin{equation}
  \label{eq:skallebas}
  \frac{\epsilon_1}{\epsilon_2} \leq \frac{2}{2+\eta}\ ,
\end{equation}
where the equality corresponds to a vanishing central pressure.
Setting $\epsilon_1=0$ corresponds to an eigendirection in the $Q=1$ subset
and is associated with taking the limit $p_c \rightarrow \infty$ for the 
interior solutions. The quotient $p_c/\rho_c$ is a gravitational
strength parameter. The Newtonian limit corresponds to small values of this
parameter. Relativistic effects are thus most pronounced when this
parameter takes as large values as possible.
The maximal value $\frac{p_c}{\rho_c}=\frac{1}{\eta-1}$, 
describing the high pressure limit, is obtained
when $\epsilon_1=0$, which corresponds to the $Q=1$ subset. 
The $Q=1$ subset can thus be expected to play a role when probing 
relativistic effects.

Solutions describing stars correspond to orbits
that start from the equilibrium point $P_3^+$ and satisfy the 
positive pressure criteria in (\ref{eq:skallebas}). 
These orbits eventually pass through the surface of
vanishing pressure at an interior point $(Q_*, S_*, C_*)$ of the state
space. From (\ref{eq:linRexp}) it follows that the linear equation of state
with $\rho_0 \neq 0$ leads to star models with finite radii.
This is expected, since star models always are
finite when $\rho\neq0$ as $p\rightarrow 0$ (see Rendall \& Schmidt 
\cite{art:RendallSchmidt1991}).      

\begin{figure}[ht]
  \begin{center}
    \epsfig{figure=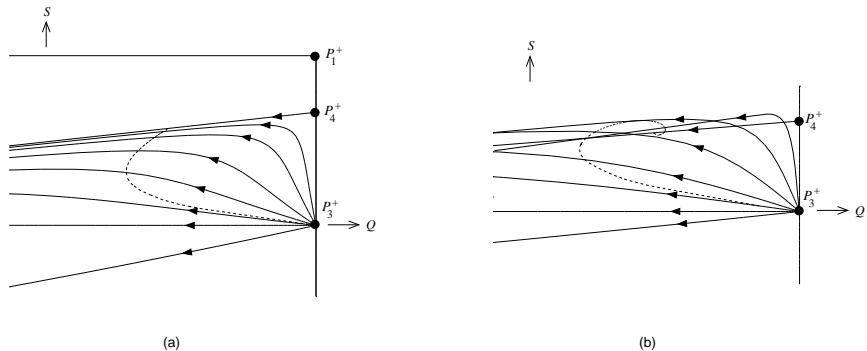, width=\textwidth}
    \caption{Orbits belonging to the regular subset for models with a
      linear equation of state, using the variables
      $\left\{Q,S,C\right\}$, projected onto 
      the plane-symmetric boundary $C=0$ for the cases (a)
      $1 \leq \eta \leq \sqrt{64/7} - 2$, and
      (b) $\eta \geq 2$. The intersection of the regular subset and
      the surface of vanishing pressure is indicated by the dotted
      lines.} 
    \label{fig:cylspace}
  \end{center}
\end{figure}

The qualitative behavior of the regular subset, projected onto the
plane-symmetric boundary ($C=0$) is shown in figure \ref{fig:cylspace}a for 
$1 \leq \eta \leq \sqrt{64/7}-2$ and in figure \ref{fig:cylspace}b for
models 
with $\eta\geq2$. The intersection between the orbits in the regular
subset and the surface of vanishing pressure is indicated by the
dashed lines. Two orbits constitute the ``high pressure''
boundary of the regular subset. The first orbit belongs to the boundary $Q=1$
and starts at $P_3^+$ and ends at $P_4^+$. This orbit
corresponds to a unique regular scale-invariant solution with
infinite radius, in contrast to models with $\rho_0\neq0$ 
(see, for example, Collins \cite{art:Collins1985}). 
Thus solutions with high central pressure are approximately described by
this solution near their center. The other orbit enters the interior state
space from the ``Tolman'' equilibrium point $P_4^+$. This orbit represents a
non-regular solution, but nevertheless plays an important role for
regular models, particularly those with high central energy density
and pressure. We will refer to this orbit as the Tolman orbit (although
a corresponding exact solution is known only for the ``stiff'' fluid case, 
$\eta=2$, see Tolman \cite{art:Tolman1939}). 

For the incompressible fluid case and low values of $\eta$, 
this orbit forms the boundary of a ``simple'' surface formed by the 
regular subset. For larger values of $\eta$ ($\eta > \sqrt{64/7} - 2$),
the self-similar Tolman point $P_4^+$ is a focus. 
This leads to a more complicated situation where ``high central pressure''
orbits in the regular subset spiral around the Tolman orbit, which
in this case also acts as a ``skeleton'' orbit for the
regular subset. 
Thus the ``regular boundary'' orbit, the Tolman point $P_4^+$,
and the Tolman orbit
are seen to play key roles for the regular subset. Moreover, it is
possible to obtain approximations for solutions with high central pressure
by piecewise joining perturbations of the regular boundary orbit and the 
Tolman orbit.

In the interior of the star, a mass function $m(r)$ can be
defined (see, for example, Misner \& Sharp
\cite{art:MisnerSharp1964}). The quotient  
$m(r)/r$ is dimensionless, and can be written as 
\begin{equation}
  \label{eq:massfunlin}
  \frac{m}{r} = \frac{C-(Q-S)^2}{2C}\ .
\end{equation}
Equation (\ref{eq:linRexp}) and (\ref{eq:massfunlin}) imply that the total 
mass $M$ of the star is given by 
\begin{equation}
  \label{eq:mrlineq}
  M = \frac{S_*\sqrt{\eta(1-Q_*^2)}}
           {(Q_*+S_*)\sqrt{8\pi\rho_0(Q_*^2-S_*^2)}}\ .
\end{equation}
This constant is to be identified with the mass parameter of the
Schwarzschild solution (\ref{eq:schwa}) when an interior solution is matched
with the exterior Schwarzschild solution.

Besides the orbits of the regular subset and the single orbit
starting from the equilibrium point $P_4^+$, there is also a
2-parameter set of orbits starting from the point $P_2^+$
and a 1-parameter set of orbits starting from $P_5^+$ (when $\eta<2$). These
two latter sets of orbits, however, describe solutions that start out with
negative mass. Positive mass is subsequently added and the solutions 
eventually aquire a total positive mass at a sufficiently large radius.

\section{Comments on possible generalizations of the equation state}
\label{sec:general}

In this Section we address
the issue whether it is possible to modify the above
formulation so as to be useful for studying models with other
barotropic equations of state. As will be seen in this series of papers,
the behavior when $\rho,p\rightarrow\infty$ 
and when $p \rightarrow 0$ constitute key  
ingredients when one attempts to find useful formulations for a given 
equation of state. The latter limit naturally leads to two types of equations 
of state; those for which $\rho \rightarrow 0$ when $p\rightarrow 0$
and those for which $\rho \rightarrow \rho_0 > 0$ when $p\rightarrow 0$. 
As shown by Rendall \& Schmidt \cite{art:RendallSchmidt1991}, the latter case
always leads to models with finite radii. In the first case the situation is 
more complicated and one sometimes have models with finite radii and 
sometimes not. 

The present formulation
can be modified to cover the situation
when $\rho \rightarrow \rho_0 > 0,  as p\rightarrow 0$, and for some 
equations of  
state this may lead to a useful approach.
Let us consider the following class of equations of state:
\begin{equation}\label{eq:lingen}
 \rho = \rho_0 + [\eta(p/\rho_0) - 1]p\ ,
\end{equation}
where $\rho_0>0$ and where $\eta$ is an analytic function satisfying
$1\leq \eta < \infty$, for all non-negative values of the pressure. Thus, 
$\eta$ is viewed as function and not a constant. Equation (\ref{eq:pdeflin}) 
yields
\begin{equation}\label{eq:etagen}
 (p/\rho_0)\eta(p/\rho_0) = \frac{Q^2 - S^2 - C}{1 - Q^2}\ ,
\end{equation}
and hence
\begin{equation}
 \eta = \eta\left(\frac{Q^2 - S^2 - C}{1 - Q^2}\right)\ .
\end{equation}
To obtain a dynamical system, 
one first writes the given equation of state in the
form (\ref{eq:lingen}). Then one uses (\ref{eq:etagen}) to directly or
indirectly determine the function $\eta$ in terms of the variables $Q,S$ 
and $C$.
Finally one introduces
$\eta(\tfrac{Q^2 - S^2 - C}{1 - Q^2})$ into the dynamical system
(\ref{eq:lineqa})-(\ref{eq:lineqc}), possibly changing the
independent variable in a suitable way. 

This formulation, however, cannot be used
when $\rho \rightarrow 0, p\rightarrow 0$, which happens
for, {\em e.g.}, polytropic equations of state. In a sequel to this paper
we will discuss other types of formulations, suitable for treating
such problems. Moreover, one can also use these formulations 
when $\rho \rightarrow \rho_0 > 0, p\rightarrow 0$ and treat models with 
for example, linear equations of state. The present formulation has
some advantages in this case, however, and it also sheds light on these other 
formulations. It may also be useful to combine the various approaches
for certain equations of state.

\section{Conclusion}
\label{sec:conclude}
We have expressed the field equations as a 3-dimensional regular
system on a compact state space for static spherically symmetric models 
with a linear equation of state. In this formulation, the existence of 
regular solutions is trivial since they all start from a hyperbolic 
equilibrium point. The fact that all models of this type have finite
radii has been naturally incorporated into the formalism. 
We have shown that all models are asymptotically self-similar for small
radii. We have obtained a global 
picture of the solution space and this has revealed that certain 
solutions, the regular scale-invariant solution, the self-similar
Tolman solution, and the non-regular solution associated with the
central infinite pressure limit, play key roles for 
understanding the solution structure, and the structure of the 
regular solutions with large central pressure in particular. 

These special solutions exist in all models with linear equations of state,
including the incompressible fluid case, 
and basically play the same role, although there are some differences
depending on if the Tolman equilibrium point is a focus or not. 
For the incompressible fluid, these solutions are intimately connected with 
the Buchdahl inequalities
(one of the inequalities is an equality for the solution 
corresponding to the incompressible Tolman orbit), see, for example, Buchdahl
\cite{art:Buchdahl1959} and Hartle \cite{art:Hartle1978}.
In a sequel to this paper we will show that similar key solutions exist 
for other equations of state as well. Thus many of the features encountered in 
this paper are typical for large classes of equations of state.
 
An interesting application of formulations of this type,
for a given equation of state, is to probe how physical features depend
on, for example, the central pressure. In addition, the understanding of the 
solution space
for a class of equation of state allows one to investigate
how different physical features, like, for example, stability properties, 
depend on the equation of state. Another possible application is
perturbation theory. Since one has a good understanding about the background 
solutions, one might investigate how details of the equation of state affect 
the possible gravitational wave forms

\begin{acknowledgment}
This research was supported by G{\aa}l\"ostiftelsen (USN), Svenska
Institutet (USN), Stiftelsen Blanceflor (USN), the University of
Waterloo (USN), and the Swedish Natural Research Council (CU).
\end{acknowledgment}

\end{article}


\begin{thebibliography}{10}

\bibitem{art:Buchdahl1959}
H.~Buchdahl, {\em Phys. Rev.} {\bf 116} (1959), 1027.

\bibitem{book:MTW1973}
K.~S.~Thorne, C.~W.~Misner, and J.~A. Wheeler, ``Gravitation'', Freeman, 
San Francisco, 1973.

\bibitem{art:Collins1985}
C.~B. Collins, {\em J. Math. Phys.} {\bf 26} (1985), 2268

\bibitem{art:DelgatyLake1998}
M.~S.~R. Delgaty and K.~Lake, {\em Comput. Phys. Commun.} {\bf 115} 
(1998), 395.

\bibitem{art:Goliathetal1998b}
J.~M. Goliath, U.~S. Nilsson, and C.~Uggla, {\em Class. Quant. Grav.} 
{\bf 15} (1998), 2841.

\bibitem{art:Hartle1978}
J.~B. Hartle, {\em Phys. Rep.} {\bf 46} (1978), 201.

\bibitem{art:Kasner1925}
E.~Kasner, {\em Trans. Amer. Math. Soc.} {\bf 27} (1925), 155.

\bibitem{art:MisnerSharp1964}
C.~W. Misner and D.~H. Sharp, {\em Phys. Rev.} {\bf 136} (1964), B571.

\bibitem{art:MisnerZapolsky1964}
C.~W. Misner and H.~S. Zapolsky, {\em Phys. Rev. Lett.} {\bf 12} (1964), 635. 

\bibitem{art:Nilssonetal1998sta}
U.~S. Nilsson, C.~Uggla, and M.~Marklund,
{\em J. Math. Phys.} {\bf  39} (1998), 3336.

\bibitem{art:RendallSchmidt1991}
A.~D. Rendall and B.~G. Schmidt,
{\em Class. Quant. Grav.} {\bf 8} (1991), 985.

\bibitem{art:Schwarzschild1916}
K.~Schwarzschild, 
{\em Sitzber. Deut. Akad. Wiss. Berlin, Kl. Math.-Phys. Tech}, page
  189, 1916.

\bibitem{book:Stephani1982}
H.~Stephani, ``General Relativity'', 
{C}ambridge {U}niversity {P}ress, Cambridge, 1982.

\bibitem{art:stergioulas}
N.~Stergioulas. Living {R}eviews in {R}elativity 1988-8. Available on the net:
http://www.livingreviews.org/Articles/Volume1/1998-8stergio, 1998.

\bibitem{art:Tolman1939}
R.~C. Tolman, {\em Phys. Rev.} {\bf 55} (1939), 364.

\bibitem{art:UgglaSurmulen1990}
C.~Uggla and H.~von-{Z}ur M{\"u}hlen,
{\em Class. Quant. Grav.} {\bf 7} (1990), 1365.

\bibitem{book:WainwrightEllis1997}
J.~Wainwright and G.~F.~R. Ellis,
``Dynamical systems in cosmology'', 
{C}ambridge {U}niversity {P}ress, Cambridge, 1997.

\bibitem{book:Wald1984}
R.~M. Wald, 
``General relativity'', 
{U}niversity of {C}hicago {P}ress, Chicago, 1984.

\end{thebibliography}
\end{document}